\documentclass[twocolumn,showpacs,preprintnumbers,amsmath,amssymb]{revtex4}
\usepackage{graphicx}
\usepackage{dcolumn}
\usepackage{bm}

\begin{document}

\title{Proton Spectroscopic Factor in $^7$Li from $^{2}$H($^6$He,$^7$Li)$n$}
\author{Z. H. Li}\email{zhli@ciae.ac.cn}
\author{E. T. Li}
\author{B. Guo }
\author{X. X. Bai}
\author{Y. J. Li}
\author{S. Q. Yan}
\author{Y. B. Wang}
\author{G. Lian}
\author{J. Su}
\author{B. X. Wang}
\author{S. Zeng}
\author{X. Fang}
\author{W. P. Liu}

\affiliation{%
China Institute of Atomic Energy, P. O. Box 275(46),  Beijing 102
413, P. R. China}%


\begin{abstract}
The angular distribution of the $^{2}$H($^6$He,$^7$Li)$n$ reaction
was measured with a secondary $^6$He beam of 36.4 MeV for the first
time. The proton spectroscopic factor of $^7$Li ground state was
extracted to be 0.41 $\pm$ 0.05 by the normalization of the
calculational differential cross sections with the distorted-wave
Born approximation to the experimental data. It was found that the
uncertainty of extracted spectroscopic factors from the one-nucleon
transfer reactions induced by deuteron may be reduced by
constraining the volume integrals of imaginary optical potentials.
\end{abstract}

\pacs{21.10.Jx, 25.45.Hi, 25.60.Je, 25.70.Hi}
\maketitle
\section{Introduction}
The essential constituents of nuclear shell model are the single
particle orbits of the mean field which are occupied by protons and
neutrons under Pauli principle. The spectroscopic factor describes
the overlap between the initial and final states and yields the
information on the occupancy of a given single particle orbit, which
plays an important role in a variety of topics on nuclear reaction
and nuclear astrophysics. Single nucleon transfer reactions such as
$(d,p)$ or $(d,n)$ have been used extensively to extract the
spectroscopic information of the single nucleon orbits in nuclei
located at or near the stability line \cite{Mac60,Ass04,Tsa05}. The
spectroscopic study of exotic nuclei becomes feasible since the
production of radioactive ion beams \cite{Zli05,Wuo05,Zli06}. These
measurements allow the extraction of the spectroscopic factors by
normalizing the calculational differential cross sections with the
distorted-wave Born approximation (DWBA) to the experimental ones at
forward angles.

The ($^7$Li,$^6$He) reaction is a valuable spectroscopic tool in the
study of nuclear reactions because the shape of its angular
distribution can be well reproduced by the DWBA calculations
\cite{Kem74}. In the calculations of ($^7$Li,$^6$He) reactions
\cite{Kem74,Whi74,Kim81,Kem77,Wil75}, the spectroscopic factor of
$^7$Li ground state was taken to be 0.59 given by Cohen and Kurath
\cite{Coh67}. F. P. Brady et al. \cite {Bra77} extracted the
spectroscopic factor of $^7$Li ground state to be S($p_{3/2}$) =
0.62 from the $^7$Li($n,d$)$^6$He reaction with 56.3 MeV neutrons.
L. Lapik\'{a}s et al. \cite{Lap99} deduced the proton spectroscopic
factor in $^7$Li to be 0.42 $\pm$ 0.04 via the measurement of the
$^7$Li($e,e^{\prime}p$) reaction. This value is 32\% smaller than
that from the $^7$Li($n,d$)$^6$He reaction. Thus, further
measurement of the $^7$Li spectroscopic factor is highly desired.

 In the present work, the angular distribution of the $^{2}$H($^6$He,$^7$Li)$n$ reaction
was measured by using a secondary $^6$He beam of 36.4 MeV, and
analyzed with DWBA. The proton spectroscopic factor in $^7$Li was
then
 extracted and compared with the existing ones.

\section{Measurement of the angular distribution}
The experiment was carried out using the secondary beam facility
\cite{Bai95, Liu03} of the HI-13 tandem accelerator, Beijing. A 46
MeV $^{7}$Li primary beam from the tandem impinged on a 4.8 cm long
deuterium gas cell at a pressure of about 1.5 atm. The front and
rear windows of the gas cell are Havar foils, each in thickness of
1.9 mg/cm$^2$. The $^{6}$He ions were produced via the
$^{2}$H($^{7}$Li, $^{6}$He)$^3$He reaction. After the magnetic
separation and focus with a dipole and a quadruple doublet, a 37.6
MeV $^{6}$He secondary beam was delivered and then collimated with a
$\phi$7- $\phi$5 mm collimator complex. The $^{6}$He beam was then
recorded by a 23 $\mu$m  thick silicon $\Delta E$ detector, which
served as both particle identification and beam normalization. The
typical purity and intensity of the $^6$He beam are 99\% and 3000
pps. The main contaminants were $^7$Li ions out of Rutherford
scattering of the primary beam in the gas cell windows as well as on
the beam tube.

\begin{figure}[h]
\includegraphics[height=5.5 cm]{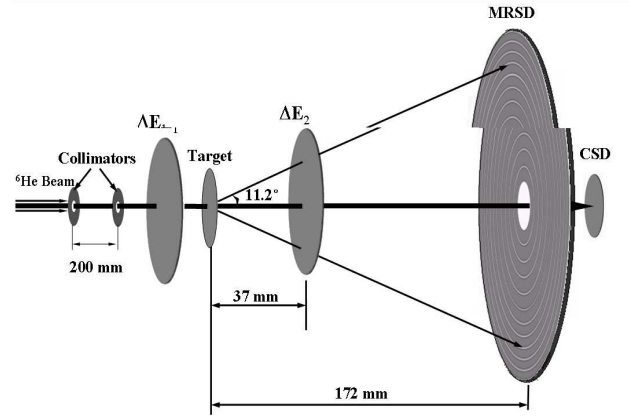}
\caption{\label{fig:setup}Schematic layout of the experimental
setup}
\end{figure}

The experimental setup is shown in Fig.~\ref{fig:setup}. A
(CD$_{2})_{n}$ foil and a carbon foil, both in the thickness of 1.7
mg/cm$^2$, were used as the targets to measure the
$^{2}$H($^6$He,$^7$Li)$n$ reaction and background, respectively. The
energy of $^6$He ions at the middle of the (CD$_{2})_{n}$ was 36.4
MeV. A 300 $\mu$m thick multi-ring semiconductor detector (MRSD)
with center hole was used as a residue energy ($E_r$) detector which
composed a $\Delta E - E_r$ counter telescope together with a 23
$\mu$m thick silicon $\Delta E$ detector and a 300 $\mu$m thick
center silicon detector (CSD). Such a detector configuration covered
the laboratory angular range from $0^\circ$ to $11.2^{\circ}$, and
the corresponding angular range in the center of mass frame for the
$^{2}$H($^6$He,$^7$Li)$n$ reaction was from $0^\circ$ to
$51.6^{\circ}$. Generally, the spectroscopic factor is extracted by
fitting the theoretical calculations to the experimental data at the
first peak in the angular distribution at forward angles
\cite{Liu04}, since the experimental angular distribution at the
backward angles is more sensitive to the inelastic coupling effects
and other high-order ones, which can not be well described
theoretically. The DWBA calculation predicts that the first peak of
the angular distribution for the $^{2}$H($^6$He,$^7$Li)$n$ reaction
is around 20$^\circ$ in the center of mass frame, thus the present
setup is propitious to the extraction of the $^7$Li spectroscopic
factor.

\begin{figure}
\includegraphics[height=10 cm]{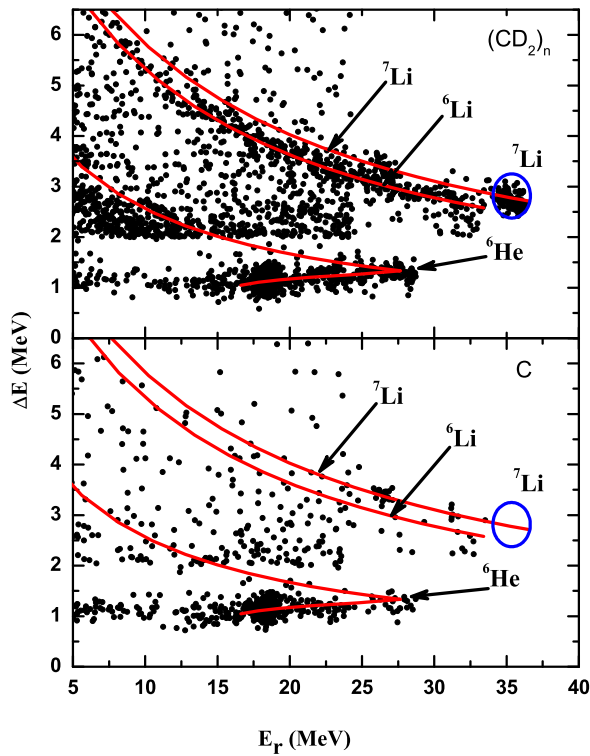}
\caption{\label{fig:fig2}(Color online) $\Delta E$ $vs.$ $E_{r}$
scatter plots of $(CD_{2})_{n}$ target (top panel) and pure carbon
target (bottom panel) measured by the fourth ring of MRSD. The red
curves are the calculational $\Delta E$ $vs.$ $E_r$ for the particle
identification of $^7$Li, $^6$Li and $^6$He. The two-dimensional
gate with blue color is the $^7$Li kinematics region from the
$^{2}$H($^6$He,$^7$Li)$n$ reaction, corresponding to the fourth
ring.}
\end{figure}

The accumulated quantity of incident $^{6}$He was approximately
$2.71\times 10^8$ for the $(CD_{2})_{n}$ target measurement, and
8.41 $\times$ $10^{7}$ for background measurement with the carbon
target. As an example, Fig.~\ref{fig:fig2} displays the $\Delta
E-E_{r}$ scatter plots of both (CD$_2$)$_n$ and carbon targets for
the fourth ring of MRSD. For the sake of saving CPU time in dealing
with the experimental data, we set a cut at $\Delta E$ = 2.0 MeV.
All the events below the cut were scaled down by a factor of 100,
while the $^{7}$Li events remain unchanged. The red curves in
Fig.~\ref{fig:fig2} are the calculated $\Delta E$ vs. $E_r$ for
$^7$Li, $^6$Li and $^6$He, respectively. The two-dimensional gate
with blue color is the $^{7}$Li kinematics region from the
$^{2}$H($^6$He,$^7$Li)$n$ reaction, corresponding to the fourth
ring. The $^{7}$Li events can be clearly identified in this figure.
We didn't find any $^{7}$Li event in the gate for the background
runs. The measured angular distribution is given in
Fig.~\ref{fig:fig3}. The uncertainties of differential cross
sections mainly arise from the statistics and the assignment of
$^{7}$Li kinematics regions.

\section{DWBA calculations}
The angular distribution measured in the this work includes the
contributions of the ground and first excited states in $^7$Li
populated by the $^{2}$H($^6$He,$^7$Li)$n$ reaction. The events of
these two states can not be separated because their energy
difference is only 0.48 MeV which is less than the energy spread
(0.62 MeV) of the $^6$He beam.

The $^{2}$H($^6$He,$^7$Li)$n$ reaction leading to the ground state
of $^7$Li is a ($3/2^-$, 1/2) $\rightarrow$ (0$^+$, 1) transition.
Parity and angular momentum considerations dictate that only
1$p_{3/2}$ pickup is possible. In the same way, the
$^{2}$H($^6$He,$^7$Li$^*$)$n$ reaction leading to the first excited
state of $^7$Li is a ($1/2^-$, 1/2) $\rightarrow$ (0$^+$, 1)
transition and only 1$p_{1/2}$ pickup contributes to the reaction.
The relationship among the experimental differential cross sections,
the DWBA calculations and the spectroscopic factors can be expressed
as
\begin{equation}\label{eq1}
({d\sigma \over d\Omega})_{exp} = S_dS_{^7Li}({d\sigma \over
d\Omega})_{gs}+S_dS_{^7Li^*}({d\sigma \over d\Omega})_{ex1},
\end{equation}
where $({d\sigma \over d\Omega})_{exp}$ is the experimental
differential cross section, $({d\sigma \over d\Omega})_{gs}$ and
$({d\sigma \over d\Omega})_{ex1}$ are the calculational differential
cross sections for the $^{2}$H($^6$He,$^7$Li)$n$ and
$^{2}$H($^6$He,$^7$Li$^*$)$n$ reactions. $S_d$ is the spectroscopic
factor for $d \rightarrow p + n$, which is derived to be 0.859 from
Ref. \cite{Blo77}. $S_{^7Li}$ and $S_{^7Li^*}$ are the proton
spectroscopic factors of the ground and first excited states in
$^7$Li. According to the translationally invariant shell model
\cite{Smi77} calculation with the code DESNA \cite{Rud85}, and
Boyarkina's wave function tables \cite{Boy73}, the ratio of
$S_{^7Li} / S_{^7Li^*}$ is 1.0 \cite{Rud05}. Thus, the proton
spectroscopic factors in $^7$Li can be extracted through Eq.
(\ref{eq1}) by the normalization of DWBA calculations to the
experimental data.

\begin{figure}
\includegraphics[height=6 cm]{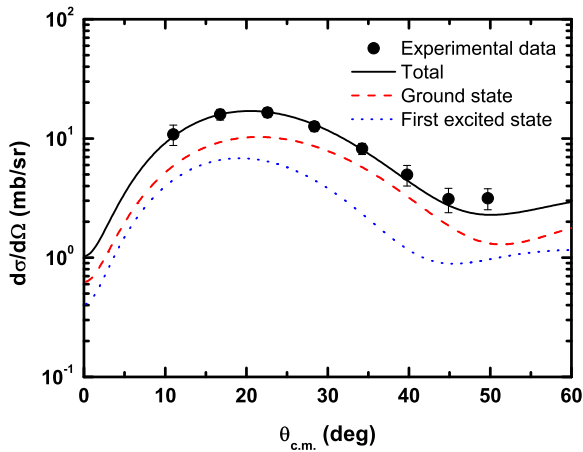}%
\caption{\label{fig:fig3}(Color online) Angular distribution of the
$^{2}$H($^6$He,$^7$Li)$n$ reaction.}
\end{figure}

The code FRESCO \cite{Tho88} was used to compute the angular
distribution of the $^{2}$H($^6$He,$^7$Li)$n$ reaction leading to
the ground and first excited states of $^7$Li. Following our
previous work in Ref. \cite{Guo07}, the effective deuteron potential
was calculated with the adiabatic model of Johnson and Soper
\cite{Sat71,Wal76}. The optical potential parameters for the
nucleon-nucleus were derived by the CH89 global systematics
\cite{Var91}. The parameterization was based on the understanding of
the optical potential theory, such as the folding model and nuclear
matter approaches instead of determining the optical potentials
phenomenologically. Up to 300 angular distributions and 9000 data
points of proton and neutron differential cross sections were
involved in the extensive database. The optical potentials derived
in this way have been successfully used in the DWBA calculations for
the $(d, p)$ reaction on light nuclei \cite{Liu04,Tsa05,Guo07}. The
optical potential parameters for the entrance and exit channels used
in our calculations, denoted as D1 and N1 respectively, are listed
in Table~\ref{tab:table1}. Five data points in the first peak of the
experimental angular distribution were used to extract the
spectroscopic factors in the DWBA calculations. The normalized
angular distributions are presented in Fig.~\ref{fig:fig3} together
with the experimental data. The dashed and dotted lines are
respectively the calculated angular distributions for
$^{2}$H($^6$He,$^7$Li)$n$ leading to the ground and first excited
states in $^7$Li, and the solid line with black color is the total
angular distribution. One can see that the shape of the experimental
angular distribution is well reproduced. $^7$Li proton spectroscopic
factor deduced from the present experimental data is 0.40 $\pm$
0.02. The error is only caused by the uncertainty of the five data
points in the first peak of the measured angular distribution.

\section{The uncertainty analysis}
\begin{figure}[h]
\includegraphics[height=6 cm]{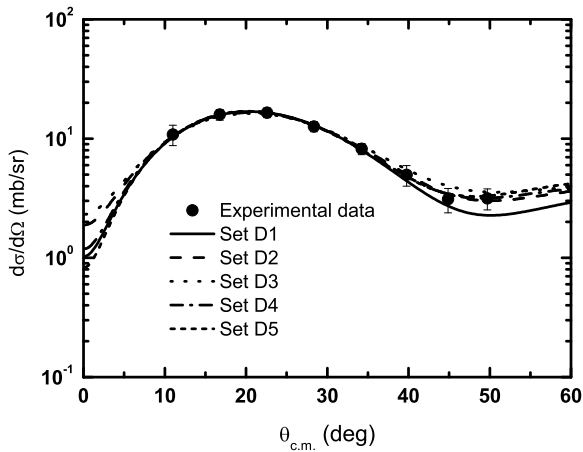}%
\caption{\label{fig:fig4}Comparison of the angular distributions of
$^{2}$H($^6$He,$^7$Li)$n$ with 5 sets of optical potential
parameters.}
\end{figure}

In general, the uncertainties of the extracted spectroscopic factor
by the DWBA calculations originate from the ambiguity of the optical
potential parameters for both the entrance and exit channels, and
that of binding potential parameters in the bound state. The optical
potential parameters for nucleon-nucleus in Refs.
\cite{Wat69,Bec69,Men71,Per76,Jeu77,Var91} were used in the present
calculations, and it is found that the potential parameters taken
from the CH89 global systematics can give a best fit in the first
peak of the $^6$He($d,n$)$^7$Li angular distribution. In order to
study the uncertainties of the spectroscopic factor associated with
the various $^6$He + $d$ optical potentials, we used the additional
four sets of $^6$He + $d$ potential parameters to extract the proton
spectroscopic factor of $^7$Li. They are labeled as D2, D3, D4 and
D5 respectively, as listed in Table~\ref{tab:table1}. Set D2 is
obtained from the analysis of an extensive set of data \cite{Dae80},
which includes the results of both polarized and unpolarized elastic
deuteron scattering on the nuclei from $^{27}$Al to $^{238}$Th in
the energy range of E$_d$ = 12 - 90 MeV. Recently, the expression
for D2 has been extrapolated to the nuclei of A $<$ 27 \cite{Liu04}.
Set D3 is based on the analysis of the elastic scattering of 52 MeV
deuterons from 27 nuclei \cite{Hin68}. Set D4 is deduced from the
elastic scattering of 30 MeV polarized deuterons from 10 nuclei
\cite{Per77}. Set D5 is the deuteron global potential for the nuclei
of Z $\geq$ 12 with deuteron energies from 12 to 25 MeV
\cite{Per76}. As a comparison, Fig.~\ref{fig:fig4} shows the angular
distributions calculated with all the five sets of the deuteron
optical potentials together with the present experimental data. One
can see that the experimental angular distribution is fairly
reproduced by all the five sets of optical potentials. The extracted
spectroscopic factors are 0.40, 0.37, 0.36, 0.45 and 0.46,
respectively. Their average is 0.41 with a standard deviation of
0.05.

\begin{table}
\caption{\label{tab:table1} Optical potential parameters used in
DWBA calculations, where $V$ , $W$ are in MeV, $r$ and $a$ are in
fm, the geometrical parameters of single particle bound state are
set to be $r_0$ = 1.25 fm and $a$ = 0.65 fm. }
\begin{ruledtabular}
\begin{tabular}{ccccccc}
Set No. & D1 & D2 & D3 & D4 & D5 & N1 \\
\hline
V   & 97.79 & 86.32 & 76.41 & 86.80 & 80.53 & 41.54 \\
$r_v$ & 1.13 & 1.17 & 1.25 & 1.13 & 1.15 & 1.41 \\
$a_v$ & 0.72 & 0.73 & 0.77 & 0.80 & 0.81 & 0.50 \\
$W_V$ & 2.05 & 0.18  \\
$W_D$ & 13.91 & 12.33 & 13.0 & 12.0 & 17.31 & 13.58 \\
$r_w$ & 1.10 & 1.325 & 1.25 & 1.56 & 1.34 & 1.35 \\
$a_w$ & 0.72 & 0.66 & 0.65 & 0.68 & 0.68 & 0.20 \\
$V_{so}$ & 5.90 & 6.98 & 6.0 & 5.2 & & 5.50 \\
$r_{so}$ & 0.68 & 1.07 & 1.25 & 0.85 & & 1.15 \\
$a_{so}$ & 0.63 & 0.66 & 0.77 & 0.48 & & 0.50 \\
$r_c$ & 1.30 & 1.30 & 1.30 & 1.30 & 1.15 \\
Ref. &\cite{Wal76} & \cite{Dae80}& \cite{Hin68} & \cite{Per77} & \cite{Per76} & \cite{Var91} \\
\end{tabular}
\end{ruledtabular}
\end{table}

In order to estimate the uncertainty of the spectroscopic factor
from the potential of $^6$He + $p$ bound state in $^7$Li, we have
tested the influence of the geometrical parameters ($r_0$ and $a$)
on the spectroscopic factor. The radius was changed from 1.10 to
1.40 fm while the diffuseness was adjusted to reproduce the rms
radius of the valence proton in $^7$Li which was calculated with the
charge rms radii of $^6$He and $^7$Li in Refs. \cite{Tan85,For09}
according to
\begin{equation}\label{eq2}
r_{^7Li}^2={1 \over Z+1}(Z r_{^6He}^2+r_p^2+{Z \over Z+1} r_v^2),
\end{equation}
where $r_{^7Li}$, $r_{^6He}$ and $r_p$ are the charge rms radii for
$^7$Li, $^6$He and proton, respectively. $r_v$ is the rms radius of
the valence proton in $^7$Li, $Z$ denotes the proton number in
$^6$He. The above changes led to a 3\% uncertainty of the deduced
spectroscopic factor, which was negligible compared with the
uncertainty from deuteron-$^6$He potential parameters. The final
value of the proton spectroscopic factor in $^7$Li is 0.41 $\pm$
0.05. The error is from the measurement (5\%) and the uncertainties
of optical potential parameters (11\%).

\begin{figure}[h]
\includegraphics[height=5.0 cm]{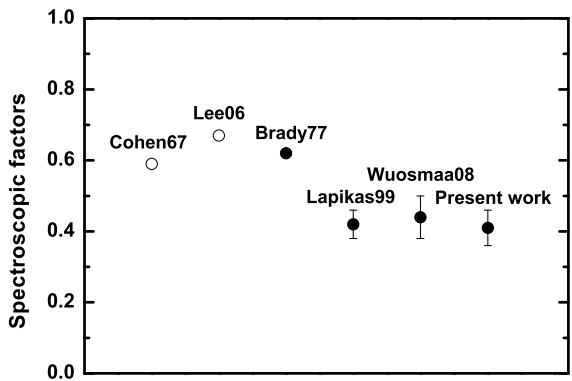}%
\caption{\label{fig:fig5}Comparison of spectroscopic factors for
$^7$Li(3/2$^-$) $\rightarrow$ $^6$He + $p$. The solid and open
circles represent the experimental and theoretical results
respectively.}
\end{figure}

Figure~\ref{fig:fig5} shows the comparison of $^7$Li spectroscopic
factors from theoretical calculations and experiments. The $^7$Li
spectroscopic factor obtained in our work is smaller than the
theoretical calculations reported in Refs. \cite{Coh67} and
\cite{Lee06}. Comparing with the experimental results, ours is 34\%
smaller than that extracted from the $^7$Li($n,d$)$^6$He reaction
\cite {Bra77}, and in good agreement with those from the
$^7$Li($e,e^{\prime}p$) reaction by L. Lapik\'{a}s et al.
\cite{Lap99} and the $^7$Li($d$,$^3$He)$^6$He reaction by A. H.
Wuosmaa et al. \cite{Wuo08} very recently.

\section{conclusion and discussion}
The $^{2}$H($^6$He,$^7$Li)$n$ angular distribution was measured with
a secondary $^6$He produced by the secondary beam facility of HI-13
tandem accelerator in Beijing. The proton spectroscopic factor in
$^7$Li ground state is extracted to be 0.41 $\pm$ 0.05, which is in
good agreement with those from the $^7$Li($e,e^{\prime}p$) reaction
\cite{Lap99} and the $^7$Li($d$,$^3$He)$^6$He reaction \cite{Wuo08}.

\begin{figure}[h]
\includegraphics[height=6 cm]{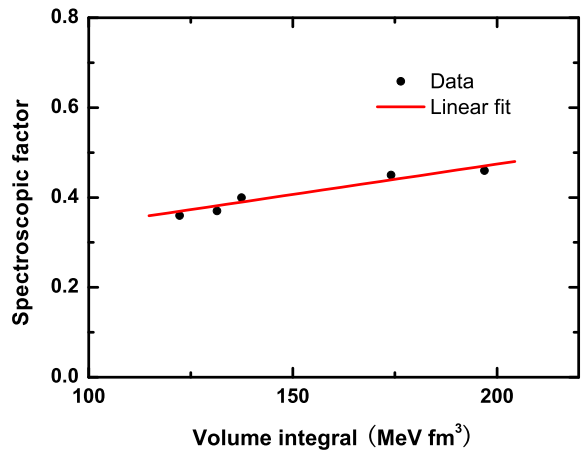}%
\caption{\label{fig:fig6}(Color online) Spectroscopic factors as a
function of the volume integral for the imaginary part of the
optical potential parameters. The red line is a linear fit of the
data points.}
\end{figure}

The error (12\%) of the spectroscopic factor given in our work
mainly arises from the uncertainty (11\%) of the optical potentials.
Thus it is important to further investigate the influence of optical
potentials. It is found that the volume integrals for the real part
of five sets potentials only differ by a factor of 3\%, while those
for the imaginary part deviate up to 20\%. The uncertainty of the
spectroscopic factor mainly arises from the uncertainty of the
imaginary optical potentials. We found a linear relationship between
the volume integrals of the imaginary part and the spectroscopic
factors, as shown in Fig.~\ref{fig:fig6}. Therefore, the uncertainty
of the extracted spectroscopic factors can be reduced by
constraining the imaginary volume integral for deuteron-nucleus
correctly. Generally speaking, the angular distribution of the
elastic scattering can provide fairly good information on the real
part of the optical potential. However, it can only give relatively
poor information on the imaginary part of the optical potential.
Consequently, it is of importance for extracting the imaginary
potential parameters to study the deuteron-nucleus reactions other
than elastic scattering.

\begin{acknowledgments}
This work is supported by the National Basic Research Programme of
China under Grant No. 2007CB815003, the National Natural Science
Foundation of China under Grant Nos. 10675173, 10705053 and
10735100.
\end{acknowledgments}

\bibliography{He6dn}
\end{document}